\documentclass[11pt,reqno]{amsart}
\usepackage{amsfonts}
\usepackage{mathrsfs}
\usepackage{amsmath}
\usepackage{amssymb,amsfonts}

\newtheorem{thm}{Theorem}[section]
\newtheorem{lem}{Lemma}[section]
\newtheorem{defn}{Definition}[section]
\newtheorem{prop}{Proposition}[section]
\numberwithin{equation}{section}
\newtheorem{rmk}{Remark}[section]

\def\pf{{\textit {Proof:} }}

\newcommand{\mysection}[1]{\section{#1}\setcounter{equation}{0}}

\newfont{\bb}{msbm10 at 12pt}

\def\hK{\hbox{\bb K}}
\def\R{\hbox{\bb R}}
\def\hC{\hbox{\bb C}}

\def\K{\mathcal K}

\newcommand{\bal}{\begin{aligned}}      \newcommand{\eal}{\end{aligned}}
\newcommand{\ba}{\begin{array}}      \newcommand{\ea}{\end{array}}
\newcommand{\bc}{\begin{center}}     \newcommand{\ec}{\end{center}}
\newcommand{\be}{\begin{enumerate}}  \newcommand{\ee}{\end{enumerate}}
\newcommand{\beq}{\begin{eqnarray}}  \newcommand{\eeq}{\end{eqnarray}}
\newcommand{\beQ}{\begin{eqnarray*}} \newcommand{\eeQ}{\end{eqnarray*}}
\newcommand{\bi}{\begin{itemize}}    \newcommand{\ei}{\end{itemize}}
\newcommand{\bt}{\begin{tabular}}    \newcommand{\et}{\end{tabular}}
\newcommand{\bdm}{\begin{displaymath}} \newcommand{\edm}{\end{displaymath}}

\def\qed{\hfill{Q.E.D.}\smallskip}

\newcommand{\ls}{\setlength{\baselineskip}{12pt}
                 \setlength{\parskip}{3mm}}

\begin{document}

\title[Positive energy theorem]{The positive energy theorem for weighted asymptotically anti-de Sitter spacetimes}

\author[Y Wang]{Yaohua Wang$^{\dag}$}
\address[]{$^{\dag}$School of Mathematics and Statistics, Henan University, Kaifeng, Henan 475004, PR China}
\email{wangyaohua@henu.edu.cn}
	
\author[X Zhang]{Xiao Zhang$^{\flat}$}
\address[]{$^{\flat}$Guangxi Center for Mathematical Research, Guangxi University, Nanning, Guangxi 530004, PR China}
\address[]{$^{\flat}$Academy of Mathematics and Systems Science, Chinese Academy of Sciences, Beijing 100190, PR China}
\address[]{$^{\flat}$School of Mathematical Sciences, University of Chinese Academy of Sciences, Beijing 100049, PR China}
\email{xzhang@gxu.edu.cn, xzhang@amss.ac.cn}

\date{}

\begin{abstract}
The positive energy theorem for weighted asymptotically flat spin manifolds was proved by Baldauf and Ozuch \cite{BO}, and for non-spin case by Chu and Zhu \cite{CZh}. In this paper, we generalize the positive energy theorem for 3-dimensional asymptotically anti-de Sitter initial data sets to weighted asymptotically anti-de Sitter initial data sets assuming that the weighted dominant energy condition holds.


\end{abstract}

\maketitle \pagenumbering{arabic}

\mysection{Introduction}\ls

The positive energy theorem plays a fundamental role in general relativity. In the case of zero cosmological constant, the positive energy theorem was first proved by Schoen and Yau \cite{SY1, SY2}, then by Witten \cite{Wi,PT}. For time-symmetric initial data sets, the dominant energy condition is equivalent to
\begin{align}\label{e1}
R \geq 0,
\end{align}
where $R$ is the scalar curvature of the initial data set. In this case, Schoen and Yau used minimal surfaces to prove that total energy is nonnegative, and zero if and only if the manifold is the Euclidean space \cite{SY1}. For non time-symmetric initial data sets, they solved the Jang equation and reduced the dominant energy condition to
\begin{equation}\label{e2}
\begin{aligned}
R-2|X|^2+2\mbox{div} X\geq 0
\end{aligned}
\end{equation}
for some falling off vector $X$, and proved that the total energy is nonnegative, and zero if and only if the manifold can be isometrically embedded into the Minkowski spacetime as a spacelike hypersurface equipped with the prescribed metric and the second fundamental form \cite{SY2}. Such strategy was also used to prove the positivity of Liu-Yau's quasi-local mass by Liu and Yau \cite{LY}. The condition (\ref{e2}) indicates that the positive energy theorem could hold true even if the scalar curvature is negative somewhere. Inspired by this observation, it was shown that the first Neumann eigenvalue is nonnegative can imply the total energy is nonnegative \cite{ZZ}. Furthermore, the second-named author derived the Schr\"{o}dinger-Lichnerowicz formula for the Dirac-Witten operator in terms of the intrinsic spin connection \cite{Z1} and used it to prove the positive total energy under the condition
\begin{equation}\label{e3}
\begin{aligned}
R+H^2-2|\nabla H|\geq 0
\end{aligned}
\end{equation}
for some falling off function $H$ \cite{Z3}. (The theorem was proved for 3-dimensional spacelike hypersurfaces, but the argument can be extended to higher dimensional spin manifolds.) Recently, the condition (\ref{e3}) was applied to prove the quantitative shielding theorem via stable $\mu$-bubbles duo to Lesourd, Unger and Yau \cite{LUY}, which was revisited via the Dirac operator by Cecchini and Zeidler \cite{CZ2} (see also \cite{CZ1}). By using the Penrose operator \cite{CZ2}, or choosing the second fundamental forms in initial data sets to be proportional to the metrics with proportional factor $-\frac{H}{n-1}$ in the dominant energy condition \cite{LLU}, Cecchini and Zeidler, as well as Lee, Lesourd and Unger showed that the quantitative shielding theorem also holds under the condition
\begin{equation}\label{e4}
\begin{aligned}
R+\frac{n}{n-1}H^2-2|\nabla H|\geq 0.
\end{aligned}
\end{equation}

In \cite{BO}, Baldauf and Ozuch defined the weighted total energy for weighted asymptotically flat manifolds
\begin{align*}
m_f(g)= m(g) + 2 \lim_{R\longrightarrow \infty}\int_{S_R}\langle\nabla f, \nu\rangle
\end{align*}
for some real function. In the case of spin manifolds, they perturbed the Dirac operator and the Laplacian operator
\begin{align*}
D_f=D-\frac{1}{2} \nabla f, \qquad \Delta_f=\Delta-\nabla_{\nabla f}
\end{align*}
and derived the Schr\"{o}dinger-Lichnerowicz formula
\begin{align*}
D_f^2=-\Delta_f +\frac{1}{4}\big(R+2\Delta f-|\nabla f|^2\big).
\end{align*}
Baldauf and Ozuch proved the positive energy theorem for $m_f(g)$ by using Witten's argument under the condition \cite{BO}
\begin{equation}\label{e0}
R+2\Delta f-|\nabla f|^2 \geq 0.
\end{equation}
A non-spin version was proved by Chu and Zhu \cite{CZh}.

It is clearly that (\ref{e2}) is stronger than (\ref{e0}) by setting $X=\nabla f$ and (\ref{e3}) is stronger than (\ref{e4}). They yield slightly different conclusions. The positivity of $m_f(g)$ does not conclude the positivity of $m(g)$. Under condition (\ref{e3}), that the total energy is equal to the total momentum implies that the manifold is the Euclidean space and $H$ is zero \cite{Z3}. But, under condition (\ref{e4}), it needs further assumptions on the metrics as well as the function $H$ to make it true. We refer \cite{BC1, BC2, CM, HL1, HL2, HZ, HL3} for the general discussing on the rigidity when $E=|P|$.

When the cosmological constant is negative, the positive energy theorem was first proved by Wang \cite{Wa}, and then by Chru\'sciel and Herzlich \cite{CH} for asymptotically anti-de Sitter initial data sets with the trivial second fundamental form. For the nontrivial second fundamental forms,
Chru\'{s}ciel, Maerten and Tod \cite{CMT} proved it by assuming the existence of center of AdS mass' coordinate systems \cite{M, CMT}, and  the authors and Xie finally proved it for general case \cite{WXZ}. An extension involving electromagnetic fields was proved by the first-named author and Xu \cite{WX}.

In this paper, we generalize the positive energy theorem in \cite{WXZ} to that of weighted asymptotically anti-de Sitter initial data. Let
$E_0$ be the total energy, $c_{i}$, ${c'}_{i}$ and $J_{i}$ be the total momenta provided in \cite{WXZ}. Let $f$ be certain suitable $C^2$ function and $E_0^f$ be the weighted total energy, $c_{i}^f$, ${c'_{i}}^f$ and $J_{i}^f$ be the weighted total momenta with respect to $f$, which are defined by (\ref{w-e-m}). If the weighted dominant energy condition (\ref{WDEC}) holds, then
\begin{align*}
E_0^f \geq \sqrt{L_f^2 -2V_f^2 +2 \big(\max\{A_f^4 -L_f^2 V_f^2, 0\}\big)^\frac{1}{2}},
\end{align*}
where $L_f$, $A_f$ and $V_f$ are given by (\ref{3constants}) (see Theorem \ref{pet} for detail).

The paper is organized as follows:
In Section 2, we define the weighted total energy-momenta for the weighted asymptotically anti-de Sitter initial data sets.
In Section 3, we provide the definition of the energy-momentum endomorphism.
In Section 4, we prove the positive energy theorem for the weighted asymptotically anti-de Sitter initial data sets satisfying the weighted dominant energy conditions. In appendix, we provide the restriction of the ten Killing vectors $U_{\alpha \beta}$ on the anti-de Sitter spacetime.

\mysection{Weighted total energy-momenta}\ls

Let $(N,\widetilde{g})$ be a 4-dimensional spacetime where $\widetilde{g}$ satisfies the Einstein field equations
\beQ
\widetilde{Ric}-\frac{\widetilde{R}}{2}\tilde{g}+\Lambda \tilde{g}=T \label{EinsteinEqs}
\eeQ
with negative cosmological constant
\begin{align*}
\Lambda=-3 \kappa^2
\end{align*}
for some positive constant $\kappa$. Let $M$ be a spacelike hypersurface of $N$ with the induced metric $g$ and the second fundamental form $h$.
An weighted initial data set is a quadruple $(M,g,h,e^{-f}dv)$ where $f$ is a real function. Denote
\beQ
R_f=R+2\Delta f-|\nabla f|^2
\eeQ
where $R$ is the scalar curvature of $M$. It satisfies the {\em weighted dominant energy condition} if the following inequality holds
\begin{equation}\label{WDEC}
\begin{aligned}
\mu^f\geq  |\nu^f|+\kappa|\nabla f|,
\end{aligned}
\end{equation}
where
\begin{align*}
\mu^f & =\frac{1}{2}\Big(R_f +6\kappa^2+\big(\sum h_{ii}\big)^2-\sum_{i,j}h_{ij}^{2}\Big),\\
\nu^f_i & =\nabla_jh_{ij}-\nabla_ih_{jj}-f_jh_{ij}.
\end{align*}

\begin{defn}\label{wads}
Weighted initial data set $(M,g,h,e^{-f}dv)$ is {\em weighted asymptotically anti-de Sitter} of order $\tau >0$ if\\
$(1)$ There is a compact set $K  \subset M$ such that $M
\setminus K$ is diffeomorphic to the disjoint union of a finite number of subsets (ends) $M_i$ and each $M_i$
is diffeomorphic to $\mathbb{R}^3 \setminus B _{r_i}$ with $B _{r_i}$ the closed ball of radius $r_i$;\\
$(2)$ Under this diffeomorphism, on each end the metric
$g_{ij}=g(\breve{e}_i,\breve{e}_j)$
is of the form $g_{ij}=\delta_{ij}+a_{ij}$ where $a_{ij}$ satisfies
\beQ
a_{ij}=O(e^{-\tau \kappa r}), \quad \breve{{\nabla}}_k a_{ij}=O(e^{-\tau \kappa r}), \quad \breve{\nabla}_l\breve{{\nabla}}_k a_{ij}=O(e^{-\tau \kappa r}),
\eeQ
the second fundamental form $h_{ij}=h(\breve{e} _i,\breve{e} _j)$ satisfies
\beQ
h_{ij}=O(e^{- \tau \kappa r}), \quad \breve{{\nabla}}_k h_{ij}=O(e^{-\tau \kappa r}),
\eeQ
and $f$ satisfies
\beQ
f=O(e^{- \tau \kappa r}), \quad \breve{{\nabla}}_k f=O(e^{-\tau \kappa r}), \quad \breve{\nabla}_l\breve{{\nabla}}_k f=O(e^{-\tau \kappa r});
\eeQ
$(3)$ There exists a distance function $\rho _z$ such that
\beQ
\mu e^{\kappa \rho_z}, \,\, \nu_i e^{\kappa \rho_z}, \,\, \nabla f e^{\kappa \rho_z} \in L^1(M)
\eeQ
where $\breve{{\nabla}}$ and $\{\breve{e}_i\}$ are the Levi-Civita connection and frame of the hyperbolic metric
\beQ
\breve{g}=dr^2+\frac{\sinh^2(\kappa r)}{\kappa^2}\big(d\theta^2+\sin^2\theta d \psi^2\big).
\eeQ
\end{defn}

When $f=0$, the above definition reduces to the asymptotically anti-de Sitter initial set \cite{WXZ}. The condition $(3)$ in Definition \ref{wads} is used to ensure that the weighted total energy-momenta are well-defined.

Throughout the paper, we always assume $\tau >\frac{3}{2}$. Denote
\beQ
\mathcal{E}_i=\breve{\nabla}^j g_{ij}-\breve
{{\nabla}}_itr_{\breve{g}}(g)-\kappa(a_{1i}-g_{1i}tr_{\breve{g}}(a)),\quad
\mathcal{P}_{ki}=h_{ki}-g_{ki}tr_{\breve{g}}(h).
\eeQ
Let $U_{\alpha \beta}$ be the Killing vectors provided in Appexdix.
Recall the following total energy-momenta provided in \cite{WXZ}, which were originally defined by Henneaux and Teitelboim \cite{HT}
\beq\label{e-m}
\begin{aligned}
E_0=&\frac{\kappa}{16\pi}\lim_{r\rightarrow \infty}\int_{S_r}\mathcal{E}_1 U_{40}^{(0)}\breve{\omega},\\
c_{i}=&\frac{\kappa}{16\pi}\lim_{r\rightarrow \infty}\int_{S_r}\mathcal{E}_1 U_{i4}^{(0)}\breve{\omega}
       +\frac{\kappa}{8\pi}\sum_{j=2}^{3}\lim_{r\rightarrow \infty}
       \int_{S_r}\mathcal{P}_{j1}U_{i4}^{(j)} \breve{\omega},\\
{c}'_{i}=&\frac{\kappa}{16\pi}\lim_{r\rightarrow \infty}\int_{S_r}\mathcal{E}_1 U_{i0}^{(0)}\breve{\omega}
        +\frac{\kappa}{8\pi}\sum_{j=2}^{3}\lim_{r\rightarrow\infty}
        \int_{S_r}\mathcal{P}_{j1}U_{i0}^{(j)} \breve{\omega},\\
J_{i}=&\frac{\kappa}{8\pi}\sum_{j=2}^{3}\lim_{r\rightarrow \infty}
       \int_{S_r}\mathcal{P}_{j1}V_{i}^{(j)} \breve{\omega},
\end{aligned}
\eeq
where $\breve{\omega}= \breve{e}^2\wedge \breve{e}^3$, $U_{\alpha\beta}=U_{\alpha\beta}^{(\gamma)}\breve{e}_{\gamma}$,
$\varepsilon _{ijl} V_i=U_{jl}$.

\begin{defn}\label{wem}
On each end, the weighted total energy-momenta for the 3-dimensional weighted asymptotically anti-de Sitter initial data set of order $\tau >\frac{3}{2}$ are
\beq\label{w-e-m}
\begin{aligned}
{E}_0^f=&{E}_0+\frac{1}{8\pi}\lim_{r\rightarrow \infty}\int_{S_r}\partial_r f \cosh(\kappa r)\breve{\omega}
,\\
{c}_{i}^f=&{c}_{i}+\frac{1}{8\pi}\cos(\kappa t)\lim_{r\rightarrow \infty}\int_{S_r}\partial_r f \sinh(\kappa r) n^i\breve{\omega},\\
{{c}'_{i}}^f=&{c}'_{i}-\frac{1}{8\pi}\sin(\kappa t)\lim_{r\rightarrow \infty}\int_{S_r}\partial_r f \sinh(\kappa r) n^i\breve{\omega},\\
J_{i}^f=&J_{i},
\end{aligned}
\eeq
where
\beQ
n^1=\sin\theta\cos\psi, \quad n^2=\sin\theta\sin\psi, \quad n^3=\cos \theta.
\eeQ
\end{defn}

Denote $\widetilde{\nabla}$, $\nabla$ the Levi-Civita connections as well as their lifts to the spinor bundle $\mathbb{S}$ with respect to $\tilde{g}$, $g$ respectively. For any $x\in M$, we choose local orthonormal basis $\{e_\alpha\}$, $\alpha =0,1,2,3$, such that
\beQ
\widetilde{\nabla}_{e_0}e_j(x)=0, \quad \nabla_{e_i}e_j(x)=0
\eeQ
for $i,j=1,2,3$. Let
\beQ
h_{ij}=\widetilde{g}(\widetilde{\nabla}_{e_i}e_0,e_j), \quad H=g^{ij} h_{ij}
\eeQ
be the component of the second fundamental form and its trace respectively. Then
\beQ
\widetilde{\nabla}_{e_i}e_0(x)=h_{ij}e_j, \quad  \widetilde{\nabla}_{e_i}e_j(x)=h_{ij}e_0,
\eeQ
Denote $\{e^\alpha \}$ the dual coframe of $\{e_\alpha\}$. Denote the spin connection
\beQ
\widehat{\nabla}_i=\nabla_i-\frac{1}{2}h_{ij}e_0\cdot e_j\cdot +\frac{\kappa\sqrt{-1}}{2}e_i\cdot.
\eeQ
It is straightforward that its adjoint operator with respect to $e^{-f}dv$ is
\begin{equation*}
\begin{aligned}
\widehat{\nabla}_i^*=&-\nabla_i-\frac{1}{2}h_{ij}e_0\cdot e_j\cdot
  +\frac{\kappa\sqrt{-1}}{2}e_i\cdot+e_i f  .
\end{aligned}
\end{equation*}
Define the weighted Dirac-Witten operator
\beQ
\widehat{D}_f=\sum_{i=1}^{3}e_i\cdot \widehat{\nabla}_i^f=D-\frac{1}{2}H e_0\cdot -\frac{3\kappa}{2}\sqrt{-1}
-\frac{1}{2}\nabla f\cdot.
\eeQ
Its adjoint operator with respect to $e^{-f}dv$ is
\beQ
\widehat{D}_f^*=D-\frac{1}{2}H e_0\cdot +\frac{3\kappa}{2}\sqrt{-1}-\frac{1}{2}\nabla f\cdot.
\eeQ

Now we derive the Weitzenb\"{o}ck formulas for $\widehat{D}_{f}$. Denote
\begin{align*}
\mu^f=&\frac{1}{2}\Big(R_f +6\kappa^2+\big(\sum h_{ii}\big)^2-\sum_{i,j}h_{ij}^{2}\Big),\\
\nu^f_i=&\nabla_jh_{ij}-\nabla_ih_{jj}-f_jh_{ij}.
\end{align*}

\begin{prop}
\begin{equation}\label{weitzenbock formula}
\widehat{D}_{f}^*\widehat{D}_{f}=\widehat{D}_{f}\widehat{D}_{f}^*=\widehat{\nabla}^{*}\widehat{\nabla}
+\widehat{\mathcal{R}}_{f},
\end{equation}
where
\begin{equation}\label{Rf}
\widehat{\mathcal{R}}_{f}=\frac{1}{2}\big(\mu ^f -\nu ^f _i e_0\cdot e_i\cdot -\kappa\sqrt{-1}\nabla f\cdot)\big.
\end{equation}
\end{prop}
\pf By straightforward computation, we have
\begin{equation}\label{weitz 1}
\begin{aligned}
\widehat{D}_{f}^*\widehat{D}_{f}=&\widehat{D}_{f}\widehat{D}_{f}^*=\nabla^*\nabla+\frac{1}{4}\Big(R+H^2+9\kappa^2\Big)\\
&-\frac{1}{2}\nabla_ih_{jj}e_i\cdot e_0\cdot +\frac{1}{2}(\Delta f)-\frac{1}{4}|\nabla f|^2+f_i\nabla_i.
\end{aligned}
\end{equation}
Note that
\begin{equation}\label{weitz 2}
\begin{aligned}
\widehat{\nabla}_{i}^*\widehat{\nabla}_{i}
=&-\nabla_i\nabla_i+\frac{1}{4}\Big(\sum_{i,j}h_{ij}^2+3\kappa^2\Big) +\frac{1}{2}\nabla_{i}h_{ij}e_0\cdot e_j\cdot\\
&-\frac{1}{2}f_ih_{ij}e_0\cdot e_j\cdot+\frac{\kappa\sqrt{-1}}{2}\nabla f\cdot+f_i\nabla_i.
\end{aligned}
\end{equation}
Combining (\ref{weitz 1}) and (\ref{weitz 2}), we have
\begin{equation*}
\begin{aligned}
\widehat{D}_f^*\widehat{D}_f=\widehat{D}_{f}\widehat{D}_{f}^*&=\widehat{\nabla}^{*}\widehat{\nabla}+\widehat{\mathcal{R}}_f.
\end{aligned}
\end{equation*}
\qed

\mysection{Energy-momentum endomorphism }\ls

The spinor bundle is $\hC ^4$ over the anti-de Sitter spacetime and the anti-de Sitter spacetime is characterized by imaginary Killing spinors satisfying the following equations
\beQ
\widetilde{\nabla}^{AdS}_X \Phi_0+\frac{\kappa \sqrt{-1}}{2}X\cdot\Phi_0=0.
\eeQ
In \cite{WXZ}, the Clifford representation
\begin{eqnarray}
\begin{aligned}
\breve{e}_0 \mapsto \begin{pmatrix}\ &\ &1 &\ \\ \ &\ & \ &1\\1&\ &\ &\ \\ \ &1&\ &\ \end{pmatrix}, &\quad
\breve{e}_1 \mapsto \begin{pmatrix}\ &\ &-1 &\ \\ \ &\ & \ &1\\1&\ &\ &\ \\ \ &-1&\ &\ \end{pmatrix},\\
\breve{e}_2 \mapsto \begin{pmatrix}\ &\ &\ &1 \\ \ &\ & 1 &\ \\\ &-1 &\ &\ \\ -1 &\ &\ &\ \end{pmatrix}, &\quad
\breve{e}_3 \mapsto \sqrt{-1}\begin{pmatrix}\ &\ &\ &1 \\ \ &\ & -1 &\ \\\ &-1 &\ &\ \\ 1 &\ &\ &\ \end{pmatrix}
\label{repre}
\end{aligned}
\end{eqnarray}
was fixed to solve the imaginary Killing spinor
\begin{equation}\label{ik}
\Phi_0 ^\lambda =\begin{pmatrix}u^+e^{\frac{\kappa r}{2}}+u^-e^{-\frac{\kappa r}{2}}\\v^+e^{\frac{\kappa
r}{2}}+v^-e^{-\frac{\kappa r}{2}}\\-\sqrt{-1}u^+e^{\frac{\kappa r}{2}}+\sqrt{-1}u^-e^{-\frac{\kappa r}{2}} \\
\sqrt{-1}v^+e^{\frac{\kappa r}{2}}-\sqrt{-1}v^-e^{-\frac{\kappa r}{2}}
\end{pmatrix},
\end{equation}
where
\begin{eqnarray*}\label{ik2}
\begin{aligned}
u^+ =&\Big(\lambda_1\cos\frac{\kappa t}{2}+\lambda_3\sin\frac{\kappa t}{2}\Big)
       e^{\frac{\sqrt{-1}}{2}\psi}\sin\frac{\theta}{2}\\
     &+\Big(\lambda_2\cos\frac{\kappa t}{2}+\lambda_4\sin\frac{\kappa t}{2}\Big)
       e^{\frac{-\sqrt{-1}}{2}\psi}\cos\frac{\theta}{2},\\
u^-=&\Big(-\lambda_1\sin\frac{\kappa t}{2}+\lambda_3\cos\frac{\kappa t}{2}\Big)
       e^{\frac{\sqrt{-1}}{2}\psi}\sin\frac{\theta}{2}\\
    &+\Big(-\lambda_2\sin\frac{\kappa t}{2}+\lambda_4\cos\frac{\kappa t}{2}\Big)
       e^{\frac{-\sqrt{-1}}{2}\psi}\cos\frac{\theta}{2},\\
v^+=&-\Big(-\lambda_1\sin\frac{\kappa t}{2}+\lambda_3\cos\frac{\kappa t}{2}\Big)
       e^{\frac{\sqrt{-1}}{2}\psi}\cos\frac{\theta}{2}\\
    &+\Big(-\lambda_2\sin\frac{\kappa t}{2}+\lambda_4\cos\frac{\kappa t}{2}\Big)
       e^{\frac{-\sqrt{-1}}{2}\psi}\sin\frac{\theta}{2}, \\
v^-=&-\Big(\lambda_1\cos\frac{\kappa t}{2}+\lambda_3\sin\frac{\kappa t}{2}\Big)
       e^{\frac{\sqrt{-1}}{2}\psi}\cos\frac{\theta}{2}\\
    &+\Big(\lambda_2\cos\frac{\kappa t}{2}+\lambda_4\sin\frac{\kappa t}{2}\Big)
       e^{\frac{-\sqrt{-1}}{2}\psi}\sin\frac{\theta}{2},
\end{aligned}
\end{eqnarray*}
and $\lambda_1$, $\lambda_2$, $\lambda_3$, $\lambda_4$ are four arbitrary complex numbers. (There is a typo that $\sqrt{-1}$ is missing for the Clifford representation of $\breve{e}_3$ in \cite{WXZ}. But the conclusions were derived in terms of (\ref{repre}) with $\sqrt{-1}$ in $\breve{e}_3$).

Denote $\hK$ the space of imaginary Killing spinors over the anti-de Sitter spacetime. It is a complex linear space with complex dimension $4$. There exists a one-to-one complex linear map
\beQ
\K : \hC ^4 \longrightarrow \hK.
\eeQ
For any given complex vector $\vec{\lambda}$,
\beQ
\K (\vec{\lambda})=\Phi _0 ^{\lambda}
\eeQ
is the unique Killing spinor correspondingly. Let $\{\breve{e} _\alpha \}$ and $\breve{\nabla}$ be the frame and Levi-Civita connection of anti-de Sitter metric (see Appendix) respectively. For an asymptotically anti-de Sitter initial data set,
\beQ
\begin{aligned}
\Theta_f = &\big(\breve{\nabla}^j g_{1j}-\breve{{\nabla}}_1tr_{\breve{g}}(g)+2e_1(f)\big)\mbox{Id} \\
           &+\kappa \sum _{l}(a_{l1}-g_{l1}tr_{\breve{g}}(a))\sqrt{-1}\breve{e}_l             \\
           &-2\sum _l (h_{l1}-g_{l1}tr_{\breve{g}}(h))\breve{e}_0\cdot\breve{e}_l
\end{aligned}
\eeQ
serves as an endomorphism of the spinor bundle over an end.

\begin{defn}\label{eme}
The energy-momentum endomorphism ${\bf Q^f}$ of the 3-dimensional weighted asymptotically anti-de Sitter initial data set is a complex linear map
\beQ
{\bf Q^f}: \hC ^4 \longrightarrow \hC ^4,
\eeQ
such that for any vector $\vec{\lambda} \in \hC ^4$,
\beQ
\langle \vec{\lambda}, {\bf Q^f}(\vec{\lambda})\rangle _{C} =\frac{1}{32\pi}\int _{S_\infty}\langle \Phi _0 ^\lambda, \Theta_f \cdot \Phi _0 ^\lambda \rangle \breve{\omega},
\eeQ
where $\langle \, , \rangle _{C}$ is the Hermitian inner product on $\hC ^4$, and $S_\infty$ is the 2-sphere at spatial infinity in $M$
and $\breve{\omega}$ is the reduced area form of $S_\infty$.
\end{defn}

Since $\Theta _f$ is Hermitian, ${\bf Q^f}$ is also Hermitian.
\begin{prop}\label{em}
Under the Clifford multiplication (\ref{repre}), the energy-momentum endomorphism has the following form
\beq \label{Qf}
\begin{aligned}
 {\bf Q^f} &=\begin{pmatrix}
 P_f          &     W_f \\
 \overline{W}_f^t  &    \hat{P}_f
 \end{pmatrix}, \qquad
 P_f=\begin{pmatrix}
 {E}_0^f-{c}_3^f             &     {c}_1^f-\sqrt{-1}{c}_2^f\\
 {c}_1^f+\sqrt{-1}{c}_2^f    &     {E}_0^f+{c}_3^f
 \end{pmatrix},\\
 W_f&=\begin{pmatrix}
 w_1  &     w_2 ^+\\
 w_2 ^-  &    -w_1
 \end{pmatrix}, \quad
 \hat{P}_f=\begin{pmatrix}
 {E}_0^f+{c}_3^f             &      -{c}_1^f+\sqrt{-1}{c}_2^f\\
 -{c}_1^f-\sqrt{-1}{c}_2^f   &      {E}_0^f-{c}_3^f
 \end{pmatrix},
\end{aligned}
\eeq
where
\beQ
w_1 = {c'_{3}}^f-\sqrt{-1}J_{3}^f, \quad w_2 ^\pm = -{c'_{1}}^f \pm J_{2}^f\pm \sqrt{-1}({c'_{2}}^f\pm J_{1}^f).
\eeQ
\end{prop}
\pf By (\ref{ik}), we have
\beQ
\begin{aligned}
\frac{1}{4}\int _{S_\infty}\langle \Phi _0 ^\lambda, \Theta_f \cdot \Phi _0 ^\lambda \rangle\breve{\omega}
=&\frac{1}{2}\int_{S_\infty}\mathcal{E}_1
            \big(\overline{u^+}u^++\overline{v^+}v^+\big) e^{\kappa r}\breve{\omega}\\
&+ \int_{S_\infty}\langle\nabla f, \nu \rangle
            \big(\overline{u^+}u^++\overline{v^+}v^+\big) e^{\kappa r}\breve{\omega}\\
&+\int_{S_\infty}\mathcal{P}_{21}
            \big(\overline{u^+}v^+ +\overline{v^+}u^+\big)e^{\kappa r} \breve{\omega}\\
&+\sqrt{-1}\int_{S_\infty}\mathcal{P}_{31}
            \big(\overline{u^+}v^+ -\overline{v^+}u^+\big)e^{\kappa r}\breve{\omega}.
\end{aligned}
\eeQ
Direct computations show that
\begin{equation*}
\begin{aligned}
\overline{u^+}u^+  +\overline{v^+}   v^+
=&\frac{1}{2}(\bar\lambda_1\lambda_1+\bar\lambda_2\lambda_2+\bar\lambda_3\lambda_3+\bar\lambda_4\lambda_4)\\
&+\frac{1}{2}\cos(\kappa t)\sin\theta \cos\psi(\bar\lambda_1\lambda_2+\bar\lambda_2\lambda_1-\bar\lambda_3\lambda_4-\bar\lambda_4\lambda_3)\\
&+\frac{1}{2}\sin(\kappa t)\sin\theta \cos\psi(\bar\lambda_1\lambda_4+\bar\lambda_2\lambda_3+\bar\lambda_3\lambda_2+\bar\lambda_4\lambda_1)\\
&+\frac{\sqrt{-1}}{2}\cos(\kappa t)\sin\theta \sin\psi(-\bar\lambda_1\lambda_2+\bar\lambda_2\lambda_1+\bar\lambda_3\lambda_4-\bar\lambda_4\lambda_3)\\
&+\frac{\sqrt{-1}}{2}\sin(\kappa t)\sin\theta \sin\psi(-\bar\lambda_1\lambda_4+\bar\lambda_2\lambda_3-\bar\lambda_3\lambda_2+\bar\lambda_4\lambda_1)\\
&+\frac{1}{2}\cos(\kappa t)\cos\theta(-\bar\lambda_1\lambda_1+\bar\lambda_2\lambda_2+\bar\lambda_3\lambda_3-\bar\lambda_4\lambda_4)\\
&+\frac{1}{2}\sin(\kappa t)\cos\theta(-\bar\lambda_1\lambda_3+\bar\lambda_2\lambda_4-\bar\lambda_3\lambda_1+\bar\lambda_4\lambda_2),\\
%
%
%
%
%
%
\overline{u^+}v^+  +\overline{v^+}u^+
=&\frac{1}{2}\sin\theta(-\bar\lambda_1\lambda_3+\bar\lambda_2\lambda_4-\bar\lambda_3\lambda_1+\bar\lambda_4\lambda_2)\\
&+\frac{1}{2}\cos\psi(\bar\lambda_1\lambda_4-\bar\lambda_2\lambda_3-\bar\lambda_3\lambda_2+\bar\lambda_4\lambda_1)\\
&+\frac{\sqrt{-1}}{2}\sin\psi(-\bar\lambda_1\lambda_4-\bar\lambda_2\lambda_3+\bar\lambda_3\lambda_2+\bar\lambda_4\lambda_1)\\
&+\frac{1}{2}\cos(\kappa t)\cos\theta \cos\psi(-\bar\lambda_1\lambda_4-\bar\lambda_2\lambda_3-\bar\lambda_3\lambda_2-\bar\lambda_4\lambda_1)\\
&+\frac{1}{2}\sin(\kappa t) \cos\theta \cos\psi(\bar\lambda_1\lambda_2+\bar\lambda_2\lambda_1-\bar\lambda_3\lambda_4-\bar\lambda_4\lambda_3)\\
&+\frac{\sqrt{-1}}{2}\cos(\kappa t)\cos\theta \sin\psi(\bar\lambda_1\lambda_4-\bar\lambda_2\lambda_3+\bar\lambda_3\lambda_2-\bar\lambda_4\lambda_1)\\
&+\frac{\sqrt{-1}}{2}\sin(\kappa t)\cos\theta \sin\psi(-\bar\lambda_1\lambda_2+\bar\lambda_2\lambda_1+\bar\lambda_3\lambda_4-\bar\lambda_4\lambda_3),\\
%
%
%
%
%
%
\overline{u^+}v^+ -\overline{v^+}u^+
=&\frac{1}{2}\sin\theta(-\bar\lambda_1\lambda_3+\bar\lambda_2\lambda_4+\bar\lambda_3\lambda_1-\bar\lambda_4\lambda_2)\\
&+\frac{1}{2}\cos(\kappa t)\cos\psi(\bar\lambda_1\lambda_4-\bar\lambda_2\lambda_3+\bar\lambda_3\lambda_2-\bar\lambda_4\lambda_1)\\
&+\frac{1}{2}\sin(\kappa t)\cos\psi(-\bar\lambda_1\lambda_2+\bar\lambda_2\lambda_1+\bar\lambda_3\lambda_4-\bar\lambda_4\lambda_3)\\
&+\frac{\sqrt{-1}}{2}\cos(\kappa t)\sin\psi(-\bar\lambda_1\lambda_4-\bar\lambda_2\lambda_3-\bar\lambda_3\lambda_2-\bar\lambda_4\lambda_1)\\
&+\frac{\sqrt{-1}}{2}\sin(\kappa t)\sin\psi(\bar\lambda_1\lambda_2+\bar\lambda_2\lambda_1-\bar\lambda_3\lambda_4-\bar\lambda_4\lambda_3)\\
&+\frac{1}{2}\cos\theta \cos\psi(-\bar\lambda_1\lambda_4-\bar\lambda_2\lambda_3+\bar\lambda_3\lambda_2+\bar\lambda_4\lambda_1)\\
&+\frac{\sqrt{-1}}{2}\cos\theta \sin\psi(\bar\lambda_1\lambda_4-\bar\lambda_2\lambda_3-\bar\lambda_3\lambda_2+\bar\lambda_4\lambda_1).
\end{aligned}
 \end{equation*}
Thus
 \beQ
\langle \vec{\lambda}, {\bf Q^f}(\vec{\lambda})\rangle _{C } =\big(\bar\lambda_1, \bar\lambda_2, \bar\lambda_3,\bar\lambda_4\big)
      {\bf Q^f}\big(\lambda_1, \lambda_2, \lambda_3, \lambda_4 \big)^t,
\eeQ
where ${\bf Q^f}$ is given by (\ref{Qf}). \qed

Denote ${\bf c}_f =(c_1^f, c_2^f, c_3^f)$, ${\bf c}_f' =({{c}' _1}^f, {{c}' _2}^f, {{c}' _3}^f)$, ${\bf J}_f =(J_1^f, J_2^f, J_3^f)$ and
\beq\label{3constants}
\begin{aligned}
L_f=&\big(|{\bf c}_f|^2 +|{\bf c}'_f|^2 +|{\bf J}_f|^2 \big)^{\frac{1}{2}},\\
A_f=&\big(|{\bf c}_f \times {\bf c}'_f |^2+|{\bf c}_f \times {\bf J}_f|^2 +| {\bf c}'_f\times {\bf J}_f |^2\big)^{\frac{1}{4}},\\
V_f=&\big(\varepsilon_{ijl}c_i^f c_{j}'^f J_l^f \big) ^{\frac{1}{3}},
\end{aligned}
\eeq
where $2L_f$, $2A_f^2$ and $V_f^3$ are the (normalized) length, surface area and volume of the parallelepiped spanned by
${\bf c}_f$, ${\bf c}'_f$ and ${\bf J}_f$. Clearly,
\beQ
L_f^2 \geq 3V_f^2.
\eeQ

\mysection{Weighted positive energy theorem}\ls

For weighted asymptotically anti-de Sitter initial data sets, we orthonormalize $\breve e_i$
\beQ
e_i=\breve e_i-\frac{1}{2}a_{ik}\breve e_k+o(e^{-\tau\kappa r}).
\eeQ
on ends. This yields a gauge transformation
\begin{align*}
\mathcal{A}: SO(\breve g) & \rightarrow SO(g),\\
\breve e_i & \mapsto e_i
\end{align*}
see, eg. \cite{Mi, AD, Z2, Z4} for detail. Denote the connection
\beQ
\overline{\nabla}=\mathcal{A}\circ\breve {{\nabla}}\circ\mathcal{A}^{-1}.
\eeQ

The following lemma can be found in, eg., \cite{Mi, AD, Z2, Z4}.
\begin{lem} \label{expression for energy-momentum}
 Let $(M,g,h)$ be a weighted $3$-dimensional asymptotically AdS initial data set. Then
\begin{equation}\label{asymptotical form for energy-momentum}
\sum_{j,\ j\neq i}Re\langle\phi,e_i\cdot
e_j\cdot({\nabla}_j-\overline{\nabla}_j)\phi\rangle=\frac{1}{4}(\breve{\nabla}^j
g_{ij}-\breve{{\nabla}}_i
tr_{\breve g}(g)+o(e^{-\tau\kappa r}))|\phi|^2
\end{equation}
for all $\phi \in \Gamma (\mathbb{S})$.
\end{lem}

Let $\Phi_0$ be the imaginary Killing spinors given by (\ref{ik}) on ends. We extend $\Phi_0$ to the whole $M$ smoothly. Let $\overline{\Phi}_0=\mathcal{A}\Phi_0$ be the $g$-pullback of $\Phi_0$. Denote $\widehat{\overline{\nabla}}_X=\overline{\nabla}_X+\frac{\sqrt{-1}}{2}\kappa X\cdot$. Then
\begin{equation}\label{modified ik}
\widehat{\overline\nabla}_j\overline{\Phi}_0
=\frac{\sqrt{-1}}{4}\kappa a_{jk}e_k\cdot\overline{\Phi}_0+o(e^{-\tau\kappa r})\overline{\Phi}_0.
\end{equation}

Now we seek the unique solution $\phi$ of $\widehat D_f$ such that $\phi$ is asymptotic to $\overline{\Phi}_0$ on ends.
For any compact set $K\subset M$, we denote $H^1(K, \mathbb{S})$ the completion of smooth sections of $\mathbb{S}|_{K}$ with respect to the norm
\beQ
||\phi||^2_{H^1(K,\mathbb{S})}=\int_K \left(|\phi|^2+|\widehat{\nabla}\phi|^2\right)e^{-f}dV_g.
\eeQ
Let $H^1(M, \mathbb{S})$ be the completion of compact supported smooth sections $C_{0}^{\infty}(\mathbb{S})$ with respect to the norm
\beQ
||\phi||^2_{H^1(M,\mathbb{S})}=\int_M \left(|\phi|^2+|\widehat{\nabla}\phi|^2\right)e^{-f}dV_g.
\eeQ
Then $H^1(K, \mathbb{S})$ and $H^1(M, \mathbb{S})$ are Hilbert spaces.

\begin{lem}\label{poincare inequality}
For 3-dimensional weighted asymptotically AdS initial data set, there is a constant $C>0$ such that
\begin{equation}
\int_M|\phi|^2e^{-f}dV_g\leq C\int_M|\widehat{\nabla}\phi|^2e^{-f}dV_g,
\end{equation}
for all $\phi\in H^1(M, \mathbb{S})$.
\end{lem}
\pf The proof is similar to the corresponding Poincar\'{e} inequalities in \cite{BC, WX}. As $C_{0}^{\infty}(\mathbb{S})$ is dense in $H^1(M, \mathbb{S})$, we can take $\phi\in C_{0}^{\infty}(\mathbb{S})$. The required inequality is then obtained by continuity.

Separate $M$ into two parts $K$ and $M\setminus K$, where $K$ is a compact set and $M\setminus K$ is constitute of ends. For any $\phi\in  C_{0}^{\infty}(\mathbb{S})$,
\begin{equation}
\int_K |\phi|^2dV_g\leq C\bigg(\int_K |\widehat{\nabla}\phi|^2 dV_g +\int_{\partial(M\setminus K)}|\phi|^2d\sigma\bigg),
\end{equation}
where $C$ is a positive constant. As
\beQ
C_1\leq e^{-f}\leq C_2
\eeQ
for some positive constants $C_1, C_2$, we obtain
\begin{equation}\label{poincare 1}
\int_K |\phi|^2e^{-f}dV_g\leq C\bigg(\int_K |\widehat{\nabla}\phi|^2e^{-f}dV_g +\int_{\partial(M\setminus K)}|\phi|^2d\sigma_g\bigg).
\end{equation}

On each end, the volume element
\beQ
e^{-f}dV_g \approx d\mu=e^{2\kappa r}dr dS_0,
\eeQ
where $dS_0$ is the area element of unit sphere $S^2(1)$. Thus we just need to prove the corresponding inequality for the volume element $d\mu$.
\begin{equation*}
\begin{aligned}
\int_{M_\infty}|\phi|^2e^{2\kappa r}drdS_0
=&\frac{1}{2\kappa}\int_{\mathbb{R}^3\setminus B(0.R_0)}|\phi|^2d(e^{2\kappa r})dS_0\\
=&-\frac{1}{2\kappa}\int_{\partial B(0,R_0)}|\phi|^2d\sigma
 -\frac{1}{2\kappa}\int_{\mathbb{R}^3\setminus B(0.R_0)}\partial_r|\phi|^2d\mu.
\end{aligned}
\end{equation*}
Note that
\begin{equation*}
\begin{aligned}
\bigg|\frac{1}{2\kappa}\int_{\mathbb{R}^3\setminus B(0,R_0)}\partial_r|\phi|^2d\mu\bigg|
\leq &\frac{1}{2\kappa}\int_{\mathbb{R}^3\setminus B(0,R_0)}\left(2|\langle\widehat{\nabla}\phi,\phi\rangle|+|h||\phi|^2+\kappa|\phi|^2\right)d\mu\\
\leq&\frac{1}{2}\int_{\mathbb{R}^3\setminus B(0,R_0)}|\phi|^2d\mu
  +\frac{1}{\kappa}\int_{\mathbb{R}^3\setminus B(0,R_0)}|\langle\widehat{\nabla}\phi,\phi\rangle|d\mu\\
  &+\frac{C}{2\kappa}e^{-\tau\kappa R_0}\int_{\mathbb{R}^3\setminus B(0,R_0)}|\phi|^2d\mu\\
\leq &\Big(\frac{1}{2}+\frac{\epsilon}{\kappa}+\frac{C}{2\kappa}e^{-\tau\kappa R_0}\Big)\int_{\mathbb{R}^3\setminus B(0,R_0)}|\phi|^2d\mu\\
  &+\frac{4}{\epsilon\kappa}\int_{\mathbb{R}^3\setminus B(0,R_0)}|\widehat{\nabla}\phi|^2d\mu,
\end{aligned}
\end{equation*}
we obtain
\begin{equation*}
\begin{aligned}
\int_{\mathbb{R}^3\setminus B(0,R_0)}|\phi|^2d\mu
\leq &-\frac{1}{2\kappa}\int_{\partial B(0,R_0)}|\phi|^2d\sigma+\frac{4}{\epsilon\kappa}\int_{\mathbb{R}^3\setminus B(0,R_0)}|\widehat{\nabla}\phi|^2d\mu\\
      &+\Big(\frac{1}{2}+\frac{\epsilon}{\kappa}+\frac{C}{2\kappa}e^{-\tau\kappa R_0}\Big)\int_{\mathbb{R}^3\setminus B(0,R_0)}|\phi|^2d\mu.
\end{aligned}
\end{equation*}
Choosing $\epsilon$ sufficiently small and $R_0$ sufficiently large, we get
\begin{equation}
\int_{\mathbb{R}^3\setminus B(0,R_0)}|\phi|^2d\mu
\leq -A_1\int_{\partial B(0,R_0)}|\phi|^2d\sigma
      +A_2\int_{\mathbb{R}^3\setminus B(0,R_0)}|\widehat{\nabla}\phi|^2,
\end{equation}
for some positive constants $A_1$ and $A_2$. This implies
\begin{equation}\label{poincare 2}
\begin{aligned}
\int_{\mathbb{R}^3\setminus B(0,R_0)}e^{-f}|\phi|^2dV_g
\leq & -B_1\int_{\partial B(0,R_0)}|\phi|^2d\sigma_g \\
     & +B_2\int_{\mathbb{R}^3\setminus B(0,R_0)}e^{-f}|\widehat{\nabla}\phi|^2dV_g
\end{aligned}
\end{equation}
for some positive constants $B_1$ and $B_2$. Combining (\ref{poincare 1}) and (\ref{poincare 2}), we get the desired Poincar\'{e} inequality.
\qed

\begin{lem}\label{zero}
Let $(M,g,h,e^{-f}dv)$ be a 3-dimensional weighted asymptotically AdS initial data set, possibly with boundary. Suppose spinors $\phi$, $\phi_\alpha$ are $C^1$ and satisfy
\begin{align*}
\widehat{\nabla}\phi=0,\quad \widehat{\nabla}\phi_\alpha=0
\end{align*}
for each $\alpha$. Then\\
$(1)$ If $\phi(x_0)=0$ for some $x_0\in M$, then $\phi\equiv0$.\\
$(2)$ If $\phi \in L^2(M_\infty)$ for some end $M_\infty$, then $\phi\equiv0$.\\
$(3)$ If $\{\phi_\alpha\}$ are linearly independent in some end, then they are linearly independent everywhere on $M$.
\end{lem}
\pf We adopt the argument in \cite{PT, Z4}. Since
\beQ
\widehat{\nabla}_i\phi=\nabla_i\phi-\frac{1}{2}h_{ij}e_0\cdot e_j\cdot \phi +\frac{\kappa\sqrt{-1}}{2}e_i\cdot\phi=0,
\eeQ
we obtain
\beq
|d \ln|\phi||\leq C(\kappa+|h|)\leq C_1 \Longrightarrow |\phi(x_1)|\geq |\phi(x)|e^{-C_1 d(x_1, x)} \label{2pt}
\eeq
on the complement of the zero set of $\phi$ on $M$, where $C_1$ is some positive constant $C_1$.\\
$(1)$ Taking $x_1 \rightarrow x_0$ in (\ref{2pt}), we get contradiction if there exists $x$ such that $\phi(x) \neq 0$.\\
$(2)$ If there exists $x_2$ on some end such that $|\phi(x_2)|\neq 0$, then (\ref{2pt}) gives
\beQ
|\phi(x)|\geq |\phi(x_2)| e^{C(|x_2|-|x|)}
\eeQ
on the end. This implies that $\phi$ is not $L^2$ which gives the contradiction. Thus $\phi\equiv0$ on ends which implies $\phi\equiv0$ on $M$.\\
$(3)$ It follows from (1). \qed

\begin{prop}\label{Dirac}
Let $(M,g,h,e^{-f}dv)$ be a 3-dimensional weighted asymptotically anti-de Sitter initial data set of order $\tau >\frac{3}{2}$. Suppose that the weighted dominant energy condition (\ref{WDEC}) holds, then there exists a unique spinor $\Phi_1$ in $H^1(\mathbb{S})$  such that
\begin{eqnarray}
\widehat D_f\big(\Phi_1+\overline{\Phi}_0 \big)=0. \label{WDirac}
\end{eqnarray}
\end{prop}
\pf The proof is similar to that of Lemma 4.2 in \cite{Z2}. Firstly we show the existence part. Define the bilinear form
\beQ
\mathcal{B}(\phi,\psi)=\int_M\langle\widehat{D}_f^\ast\phi,\widehat{D}_f^\ast\psi\rangle e^{-f}dV_g,
\eeQ
on $H^1(\mathbb{S})$. The weighted dominant energy condition implies that $\mathcal{B}(\cdot,\cdot)$ is coercive. Since $\widehat D_f \overline{\Phi}_0 \in L^2(\mathbb{S})$, $\widehat\nabla \overline{\Phi}_0 \in L^2(\mathbb{S})$, there exists a spinor $\phi_1 \in H^1(\mathbb{S})$ such that,
for any $\psi\in H^1(\mathbb{S})$,
\beQ
\int_M\langle\widehat D_f^\ast\phi_1,\widehat D_f^\ast\psi\rangle e^{-f}dV_g=-\int_M\langle\widehat D_f\overline{\Phi}_0,\psi\rangle e^{-f}dV_g.
\eeQ

Let $\Phi_1=\widehat D_f^\ast\phi_1$ and $\Phi=\Phi_1+\overline{\Phi}_0$. Then $\Phi_1 \in L^2(\mathbb{S})$ and
\beQ
\int_M\langle\Phi,\widehat D_f^\ast\psi\rangle e^{-f}dV_g=0
\eeQ
for any $\psi\in H^1(\mathbb{S})$.
Let $\phi_j$ be a sequence of $H^1(\mathbb{S})$ spinors
converging to $\Phi_1$ in $L^2(\mathbb{S})$. For any $\psi\in H^1(\mathbb{S})$, we get
\begin{eqnarray*}
\lim_{j\rightarrow\infty}\int_M\langle\widehat D_f(\phi_j+\overline{\Phi}_0),\psi \rangle e^{-f}dV_g =\lim_{j\rightarrow\infty}\int_M\langle(\phi_j+\overline{\Phi}_0),\widehat D_f^\ast\psi \rangle e^{-f}dV_g
=0.
\end{eqnarray*}
Thus $\widehat D_f(\phi_j+\overline{\Phi}_0)$ converges to $0$ weakly in $L^2(\mathbb{S})$, which implies $|\widehat D_f(\phi_j+\overline{\Phi}_0)|_{L^2}$ is bounded. Combing this with the fact that $\widehat D_f \overline{\Phi}_0 \in L^2(\mathbb{S})$, one gets
$|\widehat D_f\phi_j|_{L^2}$ is bounded. Therefore,  $\widehat D_f\Phi \in L^2(\mathbb{S})$ and $\widehat D_f\Phi=0$.

Now we prove the uniqueness. If there exist $\Phi_i\in H^1(\mathbb{S})$ such that
\beQ
\widehat D_f(\Phi_i+\overline{\Phi}_0)=0
\eeQ
for $i=1,2$. Let $\Psi=\Phi_1-\Phi_2$, then
\beQ
\widehat D_f\Psi=0, \quad  \Psi\in H^1(\mathbb{S}).
\eeQ
This implies that $\widehat \nabla \Psi=0$.
By Lemma $\ref{poincare inequality}$, we obtain $\Psi\equiv0$ and the proof of the proposition is
complete.
\qed

\begin{prop}\label{1}
Let $\Phi=\overline{\Phi}_0+\Phi_1$ be the solution of (\ref{WDirac}), where ${\Phi}_0$ is given by (\ref{ik}), then
\begin{equation}\label{integrated Weitz formula}
\begin{aligned}
\int_M|\widehat\nabla\Phi|^2\ast 1
 +\int_M\langle\Phi,\widehat{\mathcal{R}}\Phi\rangle\ast1
=8\pi (\bar\lambda_1, \bar\lambda_2, \bar\lambda_3, \bar\lambda_4) {\bf Q^f}
(\lambda_1, \lambda_2, \lambda_3, \lambda_4)^t.
\end{aligned}
\end{equation}
\end{prop}
\pf Integrating the Weitzenb\"{o}ck formula (\ref{weitzenbock formula}), we obtain
\begin{eqnarray}\label{integrated form of Weitz formula}
\begin{aligned}
\int_M \big(|\widehat\nabla\Phi|^2 +& \langle\Phi,\widehat{\mathcal{R}}_f\Phi\rangle\big) e^{-f}\\
=&\lim_{r\rightarrow
 \infty}\Re\int_{S_r}\langle\Phi,\sum_{ j\neq i }e_i\cdot
 e_j\cdot\widehat\nabla_j \Phi\rangle e^{-f}\ast e^i\\
& -\frac{1}{2}\lim_{r\rightarrow
 \infty}\Re\int_{S_r}\langle\Phi,e_i\cdot
 \nabla f\cdot\Phi\rangle e^{-f}\ast e^i\\
=&\ \frac{1}{4}\lim_{r\rightarrow
 \infty}\int_{S_r}(\breve{\nabla}^j
 g_{1j}-\breve
 {{\nabla}}_1tr_{\breve{g}}(g))|\Phi_0|^2 \breve{\omega}\\
\ &+\frac{1}{4}\lim_{r\rightarrow
 \infty}\int_{S_r} \kappa(a_{k1}-g_{k1}tr_{\breve{g}}(a))\langle\Phi_0,\sqrt{-1}\breve{e}_k\cdot\Phi_0
 \rangle\breve{\omega}\\
\ &-\frac{1}{2}\lim_{r\rightarrow
 \infty}\int_{S_r}(h_{k1}-g_{k1}tr_{\breve{g}}(h))\langle\Phi_0,\breve{e}_0\cdot\breve{e}_k\cdot\Phi_0\rangle
  \breve{\omega}\\&
  +\frac{1}{2}\lim_{r\rightarrow
 \infty}\int_{S_r}\langle\nabla f, \nu \rangle |\Phi_0|^2 \breve{\omega}\\
=&\frac{1}{4}\int _{S_\infty}\langle \Phi _0, \Theta_f \cdot \Phi _0  \rangle \breve{\omega}.
\end{aligned}
\end{eqnarray}
The detailed calculation for (\ref{integrated form of Weitz formula}) can be found in \cite{BO, WXZ}. From Definition \ref{eme} and Proposition \ref{em}, we know that the proposition follows.\qed

Now we can prove the weighted positive energy theorem for the weighted asymptotically anti-de Sitter initial data sets.
\begin{thm}\label{pet}
Let $(M,g,h,e^{-f}dv)$ be a $3$-dimensional smooth weighted asymptotically anti-de Sitter initial data set of order $\tau >\frac{3}{2}$. Suppose that the weighted dominant energy condition (\ref{WDEC}) holds, then, for each end
\begin{equation}
\label{ineq}
E_0^f \geq \sqrt{L_f^2 -2V_f^2 +2 \big(\max\{A_f^4 -L_f^2 V_f^2, 0\}\big)^\frac{1}{2}}.
\end{equation}
If $E_0^f=0$ for some end, then $M$ has only one end, ${\bf Q^f}=0$, and $(N,\widetilde{g})$ is anti-de Sitter along $M$.
\end{thm}
\pf We can derive ${\bf Q^f}$ is positive semidefinite by Proposition \ref{1} and the weighted dominant energy condition (\ref{WDEC}). Inequality (\ref{ineq}) can be proved similar to the proof of Theorem 4.1 in \cite{WXZ}.

If $E_0^f=0$ for some end and $M$ has other ends, we can choose $\overline{\Phi}_0=0$ on other ends and Lemma \ref{zero} (2) ensures ${\Phi}\equiv0$ which gives the contradiction. Therefore we have $M$ has only one end and ${\bf Q^f}=0$. By (\ref{integrated Weitz formula}), (\ref{WDEC}) and Lemma \ref{zero}, we know that there exists $\{\phi_\alpha\}$ which forms a basis of the spinor bundle everywhere over $M$ such that $\widehat{\nabla}\phi_\alpha=0$.
Standard argument shows that $(N,\widetilde{g})$ is anti-de Sitter along $M$ \cite{WXZ}.
\qed

The initial data set $(M,g,h,e^{-f}dv)$ has a weighted future/past trapped surface $(\Sigma,\bar{g},\bar{h})$ if
\begin{equation}\label{bound}
tr_{\bar g} (\bar h) \mp tr_{\bar g} (h|_\Sigma )-\nu (f) \geq 0,
\end{equation}
where $\Sigma $ is a connect component of boundary of $M$ equipped with the induced metric $\bar{g}$ and the second fundamental form $\bar{h}$, $\nu$ is the outward unit normal of $\Sigma$.

\begin{thm}\label{pet2}
Let $(M,g,h,e^{-f}dv)$ be a $3$-dimensional smooth weighted asymptotically anti-de Sitter initial data set of order $\tau >\frac{3}{2}$, which has finite number of weighted future/past trapped surfaces. Suppose that the weighted dominant energy condition (\ref{WDEC}) holds, then, for each end
\begin{equation}
\label{ineq2}
E_0^f \geq \sqrt{L_f^2 -2V_f^2 +2 \big(\max\{A_f^4 -L_f^2 V_f^2, 0\}\big)^\frac{1}{2}}.
\end{equation}
If $E_0^f=0$ for some end, then $M$ has only one end, ${\bf Q^f}=0$, and $(N,\widetilde{g})$ is anti-de Sitter along $M$. Moreover, over the weighted future/past trapped surfaces,
\beQ
tr_{\bar g} (\bar h) \mp tr_{\bar g} (h|_\Sigma )-\nu (f) = 0.
\eeQ
\end{thm}
\pf Let $e_3=\nu$ and $e_A$ be tangent to $\Sigma$. The boundary term involving $\Sigma $ in the Weitzenb\"{o}ck formula is
\begin{eqnarray*}
\begin{aligned}
   &\int_\Sigma \langle \Phi, e_3\cdot e_A\cdot \widehat\nabla_A \Phi \rangle e^{-f}-\frac{1}{2}\int_\Sigma \langle\Phi,e_3\cdot
 \nabla f\cdot\Phi\rangle e^{-f}\\
 = &\int_\Sigma \langle \Phi, e_3 \cdot e_A\cdot \widetilde\nabla_A \Phi \rangle e^{-f}
-\int_\Sigma \langle \Phi, \sqrt{-1}\kappa e_3\cdot \Phi \rangle e^{-f}\\
   &+\frac{1}{2}\int_\Sigma \langle\Phi, \nu(f) \Phi\rangle e^{-f}
    -\frac{1}{2}\int_\Sigma \langle\Phi, e_A(f) e_3\cdot e_A \cdot \Phi\rangle e^{-f}.
\end{aligned}
\end{eqnarray*}
By the standard argument in \cite{GHHP, WXZ}, if we take the boundary condition $e_0e_3\Phi=\pm\Phi$ on the weighted future/past trapped surface,
the boundary term involving $\Sigma $ in the Weitzenb\"{o}ck formula is
\begin{eqnarray*}
   \int_{\Sigma}\frac{1}{2}\big(- tr_{\bar g} (\bar h)\pm tr_{\bar g} (h|_\Sigma ) +\nu(f) \big)|\Phi|^2e^{-f}
\end{eqnarray*}
which is non-positive. This yields the theorem. \qed

\begin{rmk}
It is unclear whether the equality in (\ref{bound}) yields a contradiction to make $E_0^f$ always positive in Theorem \ref{pet2}.
\end{rmk}


{\footnotesize {\it Acknowledgement. This work is supported by the National Natural Science Foundation of China 12326602, the special foundation for Guangxi Ba Gui Scholars and Junwu Scholars.}

}


\mysection{Appendix: Killing vectors for the anti-de Sitter spacetime}

The anti-de Sitter spacetime can be viewed as the hyperboloid
\beQ
\eta_{\alpha\beta}y^\alpha y^\beta=-\frac{1}{\kappa ^2}
\eeQ
in $\R^{3,2}$ equipped with the metric
\beQ
\eta_{\alpha\beta}dy^\alpha dy^\beta= -(dy^0)^2+\sum^3_{i=1}(dy^i)^2-(dy^4)^2.
\eeQ
Under coordinate transformations
\beQ
y^0=\frac{\cos(\kappa t)}{\kappa} \cosh(\kappa r),\quad
y^i=\frac{1}{\kappa}\sinh(\kappa r) n^i,\quad
y^4=\frac{\sin(\kappa t)}{\kappa} \cosh(\kappa r),
\eeQ
where
\beQ
n^1=\sin\theta\cos\psi, \quad n^2=\sin\theta\sin\psi, \quad n^3=\cos \theta,
\eeQ
the induced anti-de Sitter metric is
\beQ
\widetilde{g}_{AdS}=-\cosh^2(\kappa r)dt^2+dr^2+\frac{\sinh^2(\kappa r)}{\kappa^2}\big(d\theta^2+\sin^2\theta d \psi^2\big).
\eeQ

Restricted on the hyperboloid, the rotation vectors for $\R^{3,2}$
\beQ
U_{\alpha\beta}=y_\alpha \frac{\partial}{\partial y^\beta}-y_\beta\frac{\partial}{\partial y^\alpha}
\eeQ
provide the ten Killing vectors for the anti-de Sitter spacetime
\beq\label{ten-K}
\begin{aligned}
 U_{10}=&\frac{\cos (\kappa t)}{\kappa} \Big[\sin\theta \cos\psi\frac{\partial}{\partial r}
        +\kappa \coth(\kappa r)\Big(\cos\theta\cos\psi\frac{\partial}{\partial \theta}
        -\frac{\sin\psi}{\sin \theta}\frac{\partial}{\partial\psi}\Big)\Big]\\
        &-\frac{\sin (\kappa t)}{\kappa} \tanh(\kappa r)\sin\theta \cos\psi\ \frac{\partial}{\partial t},\\
 U_{20}=&\frac{\cos (\kappa t)}{\kappa} \Big[\sin\theta \sin\psi\frac{\partial}{\partial r}
        +\kappa \coth(\kappa r)\Big(\cos\theta\sin\psi\frac{\partial}{\partial \theta}
        +\frac{\cos\psi}{\sin \theta}\frac{\partial}{\partial\psi}\Big)\Big]\\
        &-\frac{\sin (\kappa t)}{\kappa}\tanh(\kappa r)\sin\theta \sin\psi \frac{\partial}{\partial t},\\
 U_{30}=&\frac{\cos (\kappa t)}{\kappa}\Big[\cos\theta\frac{\partial}{\partial r}
        -\kappa \coth(\kappa r)\sin\theta\frac{\partial}{\partial \theta}\Big]
        -\frac{\sin (\kappa t)}{\kappa}\tanh(\kappa r)\cos\theta\frac{\partial}{\partial t},\\
 U_{40}=&\frac{1}{\kappa} \frac{\partial}{\partial t},\\
 U_{14}=&\frac{\sin (\kappa t)}{\kappa}  \Big[\sin\theta \cos\psi\frac{\partial}{\partial r}
        +\kappa \coth(\kappa r)\Big(\cos\theta\cos\psi\frac{\partial}{\partial \theta}
        -\frac{\sin\psi}{\sin\theta}\frac{\partial}{\partial \psi}\Big)\Big]\\
        &+\frac{\cos (\kappa t)}{\kappa}\tanh(\kappa r) \sin\theta \cos\psi\frac{\partial}{\partial t},\\
 U_{24}=& \frac{\sin (\kappa t)}{\kappa} \Big[\sin\theta \sin\psi\frac{\partial}{\partial r}
        +\kappa \coth(\kappa r)\Big(\cos\theta\sin\psi\frac{\partial}{\partial \theta}
        +\frac{\cos\psi}{\sin\theta}\frac{\partial}{\partial \psi}\Big)\Big]\\
        &+\frac{\cos (\kappa t)}{\kappa}\tanh(\kappa r)\sin\theta \sin\psi\frac{\partial}{\partial t},\\
 U_{34}=&\frac{\sin (\kappa t)}{\kappa}\Big[\cos\theta\frac{\partial}{\partial r}
        -\kappa \coth(\kappa r)\sin\theta\frac{\partial}{\partial \theta}\Big]
        +\frac{\cos (\kappa t)}{\kappa}\tanh(\kappa r)\cos\theta\frac{\partial}{\partial t},\\
 U_{12}=&\frac{\partial}{\partial \psi},\\
 U_{23}=&-\sin\psi\frac{\partial}{\partial\theta}
         -\frac{\cos\theta \cos \psi}{\sin \theta}\frac{\partial}{\partial \psi},\\
 U_{31}=&\cos\psi\frac{\partial}{\partial \theta}
        -\frac{\cos\theta \sin\psi}{\sin \theta}\frac{\partial}{\partial \psi}.
\end{aligned}
\eeq


\bigskip

\begin{thebibliography}{GGG}


\bibitem{AD} Andersson L, Dahl M. Scalar curvature rigidity for asymptotically locally hyperbolic
manifolds. Ann Glob Anal Geom, 1998, 16: 1-27
\bibitem{BC} Bartnik R, Chru\'{s}ciel P T. Boundary value problems for Dirac-type equations. J Reine Angew Math, 2005, 579: 13-73
\bibitem{BO} Baldauf J, Ozuch T. Spinors and mass on weighted manifolds. Commun Math Phys, 2022, 394: 1153-1172 
\bibitem{CH} Chru\'{s}ciel P T, Herzlich M. The mass of asymptotically hyperbolic Riemannian manifolds. Pacific Jour Math, 2003, 212: 231-264
\bibitem{BC1} Beig R, Chru\'{s}ciel P T. Killing vectors in asymptotically flat space-times. I. Asymptotically translational Killing vectors and the rigid positive energy theorem. J Math Phys, 1996, 37: 1939-1961
\bibitem{BC2} Beig R, Chru\'{s}ciel P T. Killing initial data. Class Quantum Grav, 1997, 14: A83-A92
\bibitem{CM} Chru\'{s}ciel P T, Maerten D. Killing vectors in asymptotically flat space-times. II.
Asymptotically translational Killing vectors and the rigid positive energy theorem in higher dimensions. J Math Phys, 2006, 47: 022502
\bibitem{CMT}  Chru\'{s}ciel P T, Maerten D, Tod P. Rigid upper bounds for the angular momentum and centre of mass of non-singular
asymptotically anti-de Sitter space-times. J High Energy Phys, 2006, 11: 084
\bibitem{CZ1} Cecchini S, Zeidler R. Scalar and mean curvature comparison via the Dirac operator.
 arXiv: 2103.06833, 2021
\bibitem{CZ2} Cecchini S, Zeidler R. The positive mass theorem and distance estimates in the spin setting. arXiv: 2108.11972, 2021
\bibitem{CZh} Chu J, Zhu J. A non-spin method to the positive weighted mass theorem for weighted manifolds. arXiv: 2305.12909, 2023
\bibitem{GHHP} Gibbons G, Hawking S, Howrowitz G, Perry M. Positive mass theorems for black holes. Commun Math Phys, {\bf 88}, 1983: 295-308
\bibitem{HZ} Hirsch S, Zhang Y. The case of equality for the spacetime positive mass theorem. J
Geom Anal, 2023, 33: No. 30
\bibitem{HL1} Huang L H, Lee D A. Equality in the spacetime positive mass theorem. Commun Math
Phys, 2020, 376: 2379-2407 
\bibitem{HL2} Huang L H, Lee D A. Bartnik mass minimizing initial data sets and improvability
of the dominant energy scalar. arXiv: 2007.00593, 2020
\bibitem{HL3} Huang L H, Lee D A. Equality in the spacetime positive mass theorem II. arXiv: 2302.06040, 2023
\bibitem{HT} Henneaux M, Teitelboim C. Asymptotically anti-de Sitter spaces. Commun Math Phys, 1985, 98: 391-424
\bibitem{LLU} Lee D A, Lesourd M, Unger R. Density and positive mass theorems for incomplete manifolds. Commun Math Phys, 2022, 395: 643-677
\bibitem{LUY} Lesourd M, Unger R, Yau S T. The positive mass theorem with arbitrary ends. arXiv: 2103.02744, 2021
\bibitem{LY} Liu C C M, Yau S T. Positivity of quasi-local mass. II. J Amer Math Soc, 2006, 19: 181-204
\bibitem{M} Maerten D. Positive energy-momentum theorem for AdS-asymptotically hyperbolic manifolds. Ann Henri Poincar\'{e}, 2006, 7: 975-1011
\bibitem{Mi} Min-Oo M. Scalar curvature rigidity of asymptotically hyperbolic spin manifolds. Math Ann, 1989, 285: 527-539
\bibitem{PT} Parker T, Taubes C. On Witten's proof of the positive energy theorem. Commun Math Phys, 1982, 84: 223-238
\bibitem{SY1} Schoen R, Yau S T. On the proof of the positive mass conjecture in general relativity. Commun Math Phys, 1979, 65: 45-76
\bibitem{SY2} Schoen R, Yau S T. Proof of the positive mass theorem. II. Commun Math Phys, 1981, 79: 231-260
\bibitem{Wa} Wang X. Mass for asymptotically hyperbolic manifolds. J Differential Geom, 2001, 57: 273-299
\bibitem{WXZ} Wang Y, Xie N, Zhang X. The positive energy theorem for asymptotically anti-de Sitter spacetimes. Commun Contemp Math, 2015, 17: 389-396
\bibitem{WX} Wang Y, Xu X. Hyperbolic positive energy theorem with electromagnetic fields. Class Quantum Grav, 2015, 32: 025007
\bibitem{Wi} Witten E. A new proof of the positive energy theorem. Commun Math Phys, 1981, 80: 381-402
\bibitem{ZZ} Zhang L, Zhang X. Remarks on positive mass theorem. Commun Math Phys, 2000, 208: 663-669
\bibitem{Z1} Zhang X. Positive mass conjecture for five-dimensional Lorentzian manifolds. J Math Phys, 1999, 40: 3540-3552
\bibitem{Z2} Zhang X. Strongly asymptotically hyperbolic Spin manifolds. Math Res Lett, 2000, 7: 719-727
\bibitem{Z3} Zhang X. Positive mass theorem for modified energy condition. in {\it Morse theory, minimax theory and their applications to nonlinear differential equations}, New Stud Adv Math, 2003, 1: 275-282
\bibitem{Z4} Zhang X. A definition of total energy-momentua and the positive mass theorem on asymptotically hyperbolic 3-manifolds I. Commun Math Phys, 2004, 249: 529-548


\end{thebibliography}
\end{document}